# SDT-DCSCN for Simultaneous Super-Resolution and Deblurring of Text Images

Hala Neji [a,b,c], Mohamed Ben Halima [b], Javier Nogueras-Iso [c], Tarek. M. Hamdani [b,g], Abdulrahman M. Qahtani [e], Omar Almutiry [f], Habib Dhahri [f], Adel M. Alimi [b,d]

[a]*National Engineering School of Gabes (ENIG), University of Gabes, Tunisia*
[b]*REGIM Lab (Research Groups in Intelligent Machines), University of Sfax,*
*National Engineering School of Sfax (ENIS), BP 1173, 3038 Sfax, Tunisia*
[c]*Aragon Institute of Engineering Research (I3A), University of Zaragoza, Spain*
[d]*Department of Electrical and Electronic Engineering Science, Faculty of Engineering and the Built Environment, University of Johannesburg, South Africa*
[e]*Department of Computer Science, College of Computers and Information Technology, Taif University, P.O.Box. 11099, Taif 21944, Saudi Arabia*
[f]*College of Applied Computer Science, King Saud University, Riyadh, Saudi Arabia.*
[g]*University of Monastir, ISIMa, BP 49, Mahdia, 5111, Tunisia.*

**ABSTRACT**

Deep convolutional neural networks (Deep CNN) have achieved hopeful performance for single image super-resolution. In particular, the Deep CNN skip Connection and Network in Network (DCSCN) architecture has been successfully applied to natural images super-resolution. In this work we propose an approach called SDT-DCSCN that jointly performs super-resolution and deblurring of low-resolution blurry text images based on DCSCN. Our approach uses subsampled blurry images in the input and original sharp images as ground truth. The used architecture is consists of a higher number of filters in the input CNN layer to a better analysis of the text details. The experimental results have achieved state-of-the-art performance in the peak-signal-to-noise ratio (PSNR), the structural similarity index measure (SSIM), the information fidelity criterion (IFC) and Visual Information Fidelity (VIF) metrics. Thus, we confirm that DCSCN provides satisfactory results for enhancement tasks on low blurry images. The quantitative and qualitative evaluation on different datasets prove the high performance of our model to reconstruct high-resolution and sharp text images with PSNR=20.406, SSIM= 0.877,VIF=0.351, IFC=2.868 for scale 4 compared to DCSCN with PSNR=15.553, SSIM= 0.621, VIF=0.166 IFC=1.129. In addition, in terms of computational time, our proposed method gives competitive performance compared to state of the art methods.

## 1. Introduction

Image quality enhancement is an essential step in image processing, , overall in the case of low quality images captured by mobile cameras or digitization devices with limited functionalities. The stored images may present problems of motion blur, out of focus blur, low-resolution (some cameras have low resolution hardware), noise (originated by bad quality of hardcopy papers or digitization mechanism). Therefore, there are several categories of image enhancement such as deblurring, denoising, dehazing and super-resolution. One of the hardest challenges in image processing is the super-resolution of blurry text images Zhang et al. (2018a,b); Liang et al. (2019) . In this paper, we are interested in improving the quality of text images to facilitate document image analysis tasks such as optical character recognition (OCR) and further text processing. In recent years, many works have focused on the super-resolution of images. For instance, Deep convolutional neural networks (Deep CNN) have achieved hopeful performance for single image super-resolution. However, there is little research in the literature about the super-resolution of blurry text images, i.e address jointly the problems of deblurring and super-resolving

*Corresponding author: Tel.: +216-25-477-540;*
Corresponding author
*e-mail:* hala.neji@ieee.org (Hala Neji), jnog@unizar.es (Javier Nogueras-Iso), tarek.hamdani@ieee.org (Tarek. M. Hamdani), amqahtani@tu.edu.sa (Abdulrahman M. Qahtani), oalmutiry@ksu.edu.sa (Omar Almutiry), hdhahri@ksu.edu.sa(Habib Dhahri), adel.alimi@ieee.org (Adel M. Alimi)

text images.

The main objective of this paper is to achieve these two joint problems of deblurring and super-resolution of text images without having knowledge of the blur kernel. Our proposal is based on the Deep CNN skip Connection and Network in Network (DCSCN)Yamanaka et al. (2017) architecture, which has been successfully applied in the past for natural images super-resolution. Within the context of our proposal, we use DCSCN to recover a sharp high resolution image from a low blurred input image without information of the blur kernel. DCSCN is a fully convolutional neural network consisting of two sections. The first one is the feature extraction network, which is in charge of extracting both the local and the global image features. The second one is the reconstruction network, which is responsible of reconstructing the image details. One of the main advantages of this model is that it provides an efficient computation for image reconstruction, achieving a computation time lower than comparable state-of-the-art approaches.

The remainder of this paper is organized as follows. We present the state-of-the-art methods in Section 2. Section 3 presents the details about the proposed method. The experiment results and discussions are described in Section 4. Section 5 provides some concluding remarks.

## 2. RELATED WORK

### 2.1. Image Super-resolution

In recent years, there have been remarkable advances in image super-resolution. We can identify two main categories of super-resolution methods: exemplar-based methods and regression-based methods.

Regarding exemplar-based methods, Wang et al Wang et al. (2012) proposed a method for learning a semi-coupled dictionary that consists of a pair of dictionaries (one for high resolution domain, and another for low resolution domain) and a mapping function. In the same line of using dictionaries, Yang et al Yang et al. (2010) trained jointly two dictionaries for low and high resolution image patches and establish a similarity function between the sparse representations of low and high resolution image patch pairs. Zeng et al. Zeng et al. (2018) also proposed a single image super-resolution method that learns iteratively non-linear mappings between low and high resolution sparse representations. Focused on the problem of generating high resolution images from face images, Yang et al Yang et al. (2013) proposed to learn statistical priors by exploiting local image structures (facial components, contours and smooth regions).

With respect to regression-based methods, we can cite the work of Zhang et al Zhang et al. (2010), who proposed a non-local kernel regression framework that takes profit of non-local self-similarity of image patches that tend to repeat themselves in natural images and the local structural regularity properties in image patches. Sun et al. Sun et al. (2008) exploited the use of gradient profiles describing the shape and sharpness of image gradients to estimate a high-resolution image from a low resolution image. In the case of Yang el al. Yang and Yang (2013), they proposed a method that learns statistical priors by applying simple functions to exemplars, which are extracted by dividing the feature space in numerous subspaces. Another relevant work is the one proposed by Timofte et al Timofte et al. (2014), who combined the simple functions approach with the use of anchored neighborhood regression, an approach that learns sparse dictionaries and regressors anchored to the dictionary atoms.

Last, we must mention the existence of recent works that have successfully applied convolutional neural networks (CNN) for single image super-resolution. For instance, Yamanaka et al. Yamanaka et al. (2017) proposed a deep CNN that has achieved hopeful performance for single image super-resolution. Jiang et al. Jiang et al. (2020) proposed a hierarchical dense connection network for image super-resolution that includes a hierarchical matrix structure providing interleaved pathways for information fusion and gradient optimization with a shallower depth, which facilitates a faster reconstruction. The work of Tran et al. Tran and Ho-Phuoc (2019) is also especially relevant as they have focused on text image super-resolution by adapting a Deep Laplacian Pyramid Network. Last, Mei et al. Mei et al. (2020) present a Cross-Scale Non-Local (CS-NL) attention module for image super-resolution, which is able to discover the widely existing cross-scale feature similarities in natural images.

### 2.2. Image deblurring

Image deblurring has been studied for a long time, and in recent years several works have specially focused on the problem of deblurring text images by means of Bayesian-based methods. For instance, Ljubenovic et al. Ljubenovic et al. (2017) proposed the use of a dictionary-based prior for class-adapted blind deblurring of document images. Jiang et al. Jiang et al. (2017) have also proposed a method based on the two-tone prior for text image deblurring.

Pan et al. Pan et al. (2016) presented a genetic approach with two main ideas: modify the prior to assume that the dark channel of natural images is sparse instead of zero, and impose the sparsity for kernel estimation. They used CNN to predict the blur kernel.

In addition, CNN networks have been also applied for text image deblurring. An example of this kind of works is the one proposed by Hradis et al. Hradiš et al. (2015), an end-to-end method to generate sharp images from blurred ones using CNN. Their network, consisting of 15 convolutional layers, is trained on 3 classes (blur, sharp image and kernel blur). Last, Neji et al. Neji et al. (2021) proposes Blur2Sharp CycleGAN, an end-to-end model for image text deblurring using cycle-consistent adversarial networks that generates a sharp image from a blurry one and shows how Cycle-Consistent Generative Adversarial Networks (CycleGAN) can be used in document deblurring.

### 2.3. Joint super-resolution and deblurring

There have been recent proposals to solve jointly the image super-resolution and the deblurring problems using deep learning.

Zhang et al. Zhang et al. (2018a) suggested a deep encoder-decoder network (ED-DSRN) to solve the problem of blurry images degraded by Gaussian blur kernel. Zhang, Dong et al.

Zhang et al. (2018b) proposed a method for weighting several enhanced versions of an input image with the weights predicted by a CNN. Liang et al. Liang et al. (2019) proposed a novel dual supervised network (DSN) to solve the deblurring and super-resolution problems.

Du et al. Du et al. (2019) suggested an approach based on CNN to reconstruct high resolution images from low-blurry ones. Liu et al. Liu et al. (2020) proposed a deep decoupled cooperative learning based CNN deblurring model to achieve disentangling and synthesizing single image super-resolution and motion deblurring. Lumentut et al. Lumentut and Park (2020) proposed a framework for the light field (LF) image enhancement using a deep neural net to super-solve the LF spatial deblurring and super-resolution under 6-degree-of-freedom camera motion. Albluwi et al. Albluwi et al. (2019) also proposed a method for single image super-resolution to tackle blur with the down-sampling of images by using CNN. Last, Quan et al Quan et al. (2020) proposed a CNN-based method for joint deblurring and super-resolution on degraded text images that introduces a collaborative mechanism to detect the relationship between the local measurement of high-frequency components and the full frequency spectrum of images. Making profit from Generative Adversarial Networks (GAN), several works have been proposed in recent years. Yun and Park Yun et al. (2020) used GAN to reconstruct high-resolution facial images by simultaneously generating a high-resolution image with and without blur. Li et al. Li et al. (2019) proposed a novel approach using GAN with Pixel and Perceptual Regularizations, denoted P2GAN, to restore single motion blurry and low-resolution images jointly into clear and high-resolution images. Du et al. Du et al. (2019) also proposed a method based on GAN for reconstructing clear high-resolution images directly from blurred low-resolution natural scene images. And focusing on the context of images combining faces and text, Xu et al.Xu et al. (2017) trained a GAN network for super-resolving this kind of images.

## 3. PROPOSED METHOD

As mentioned in the introduction, DCSCN Yamanaka et al. (2017) is a deep neural network oriented to super-solve scene images. This model is a fully convolutional neural network that contains two successive blocks: a feature extraction network and an image reconstruction network. Figure 1 shows our proposed adjustment of the original DCSCN architecture for the joint task of super-resolution and deblurring. Starting from scratch with the original architecture for training the model, we have adjusted the network to optimize the results on text images by changing the network architecture and the hyper-parameters (see table 1). The following subsections explain the details of the component networks and the two proposed versions of our method: ST-DCSCN and SDT-DCSCN.

### 3.1. Feature extraction

Many deep learning methods have a pre-processing step which is the up-sampling of the images input. In these methods the single image super-resolution networks can be pixel-wise, which is the main reason to have a big network with a large CNN rather than the complexity of computation. The fact is that these methods require multiple GPU computing to accelerate the computation, but these computation resources are not always available. To solve this issue, the DCSCN model uses a feature extraction section to decrease the size of the network and have a fast response time. DCSCN uses the original image as input of the network and the up-sampling step is inside the architecture, which decreases the number of layers and obtains a better performance with faster computation.

As an initial configuration of our architecture we started with the same network used in DCSCN, which contains seven 3x3 CNNs with *ReLU* layer activator. Since the results were promising, we tried to optimize the network for text images and tested several activators such as *ReLU*, *LeakyReLU*, *PReLU*, *Sigmoid*, *Tanh* and *SELU*. Finally, we fixed the activator at *PReLU* because it obtained the best results to train the network (see table 1).

In addition, to extract features efficiently, we evaluated different number of layers. On the one hand, we wanted to use the minimum layers to have a fast performance. On the other hand, we aimed to solve correctly the problem of deblurring and super-resolution at the same time. After many experiments, we fixed the number of filters of first feature-extraction CNNs to 196 and the number of filters of last feature-extraction CNNs to 32. Unlike to DCSCN, which has seven layers, we propose a cascade of eight 3x3 CNNs.

With respect to other parameters, we have decreased the filter decay gamma from 1.5 to 1.2. In addition, we have tested different initializers for weights: *Uniform*, *Stddev*, *Xavier*, *He*, *Identity* or *Zero*. At the end, we chose He initializer.

In summary, with this network the input image cascades to a set of CNN weights, biases and non-linear layers. After, all the extracted features are connected with skip connection to the next section of the network for reconstructing the high-resolution and sharp image.

### 3.2. Image reconstruction

In the reconstruction network, the CNN structure is parallelized in the same way as in the Network in Network architecture Lin et al. (2013). The main benefit of this structure is to reduce the dimensions of the previous layer for faster computation and to add more non-linearity to enhance the potential representation of the network. Therefore, we can lessen the number of transposed CNN filters. In addition, it must be noted that 1x1 CNNs have a computation cost 9 times lower than 3x3 CNNs. Thus, the size of the network will be reduced. As it has been mentioned before, all the global/local features are extracted and concatenated at the input layer of the reconstruction network. The input data will be very large, so it is a good idea to use 1x1 CNNs to reduce the dimensions and generate the high resolution sharp image. As shown in figure 1, the dark blue CNN generates 4 outputs channels (in case the scale factor is equals to 2) and at the end of the model the low-resolution image is reshaped to a high-resolution and sharp one by adding the bicubic up-sampled original input image. We fixed the number of reconstruction layers to 3. The number of CNN filters in A1 at the reconstruction network is 64, and the number of CNN filters in B1 and B2 at the reconstruction network is set to 32.

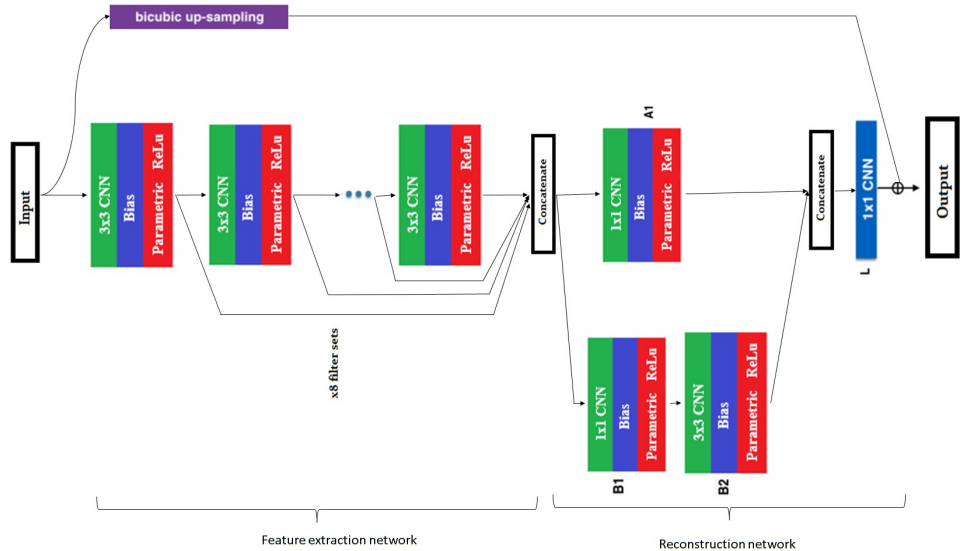

**Fig. 1.** Adjustment of DCSCN architecture proposed by Yamanaka et al Yamanaka et al. (2017).

**Table 1.** The number of filters of each CNN layer of our proposed model

| Architectural Variants | Feature extraction network | | | | | | | | Reconstruct network | | | |
|---|---|---|---|---|---|---|---|---|---|---|---|---|
| | 1 | 2 | 3 | 4 | 5 | 6 | 7 | 8 | A1 | B1 | B2 | L |
| P-Relu | 96 | 76 | 65 | 55 | 47 | 39 | 32 | | 64 | 32 | 32 | 4 |
| Relu | 196 | 163 | 138 | 115 | 93 | 72 | 51 | 32 | 64 | 32 | 32 | 4 |
| Sigmoid | 19 | 16 | 14 | 13 | 11 | 9 | 8 | 7 | 128 | 3 | 3 | 4 |
| | 128 | 103 | 83 | 66 | 49 | 33 | 18 | 3 | 19 | 7 | 7 | 4 |
| ST/SDT-DCSCN | 196 | 163 | 138 | 115 | 93 | 72 | 51 | 32 | 64 | 32 | 32 | 4 |

### 3.3. Super-Resolution of Text Images (ST-DCSCN)

We have adjusted DCSCN for super resolving text images, and we have named this adjustment as ST-DCSCN. Figure 2 shows the training process of DCSCN. The sharp images are sub-sampled with by a scale S (S=2 or S=4). These sub-sampled images, together with their bicubic up-sampling, are used as the input of the skip connection and network in network to super solve the text images. Finally, the output will be a high resolution and sharp image.

### 3.4. Simultaneous Super-Resolution and Deblurring of Text Images (SDT-DCSCN)

In this second adjusted version of DCSCN, which we call SDT-DCSCN, out hypothesis is to verify if the neural network can learn to transform blurred images into sharp images by training the network using as input the blurred images and the bicubic upsampling of the sharp images. In addition, the true sharp images are used to check the quality of the neural network output. As shown in figure 3, we use as input both the blurred images and their corresponding sharp versions for training the network.The images are subsampled with a scale S (S=2 or 4). Then,we use the low resolution and blurred images, together with the up-sampling of the sharp versions, as input of the skip connection and network in network. Finally, the output will be a deblurred image with high resolution.

## 4. EXPERIMENT RESULTS

### 4.1. Datasets and environment configuration

For training the network we used the dataset proposed by Hradis et al. Hradiš et al. (2015), which contains 66,742 paired images: blurred images and sharp images. In the first step, we used only the sharp images as input for the training of ST-DCSCN model . Then we used both of blur and sharp images as input for the training of SDT-DCSCN.

With respect to the configuration of the training network, we decided to crop randomly the input images with a patch size of 32x32, generating around 600,000 training images. For the generation of the training set, we fixed the pixel shuffler filters at 1. All the images, originally in RGB color, were converted to YCbCr images, and only the Y-channel was processed. As mentioned before, each training image was split into 32x32 patches and 20 patches were used as a mini-batch. In addition, the *He* initializer He et al. (2015) was used for each CNN, and all biases and *PReLUs* were initialized to 0. We fixed the dropout to 0.8 used in each output of *PReLU* layers. To minimize the loss, we used the Adam optimizer with a learning rate equals to 0.002. In addition, it must be noted that all the computation works were executed on an Ubuntu server with NVIDIA Quadro P6000 GPUs.

About the testing dataset, we decided to test the model on two different datasets: 100 paired images of the dataset proposed by Hradis et al.Hradiš et al. (2015), which includes both blurred and sharp low resolution images; and the dataset proposed by

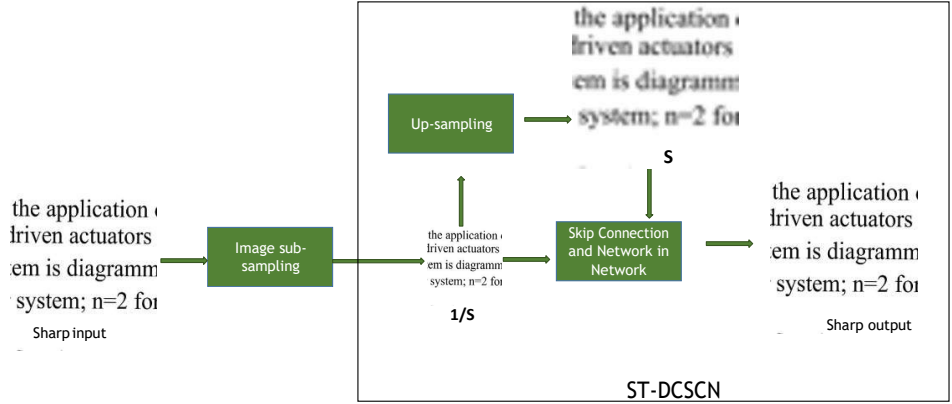

**Fig. 2. Data processing for training the super-resolution ST-DCSCN model**

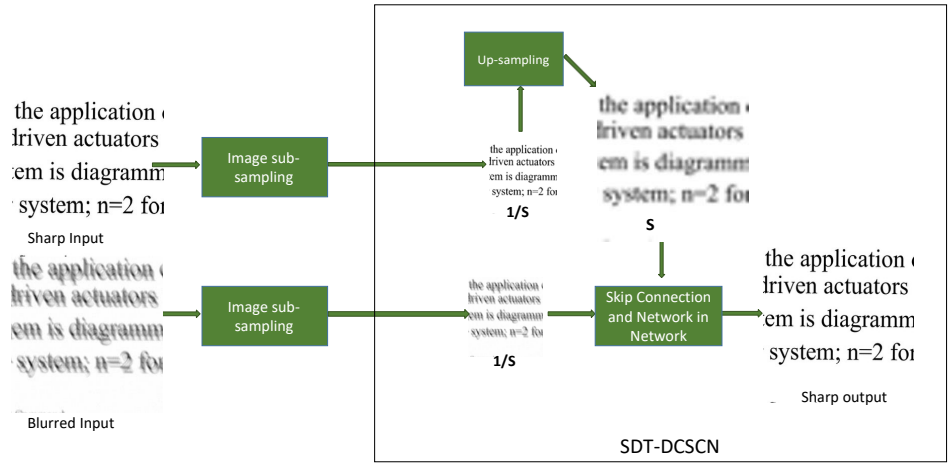

**Fig. 3. Data processing for training Simultaneous Super-Resolution and Deblurring STD-DCSCN**

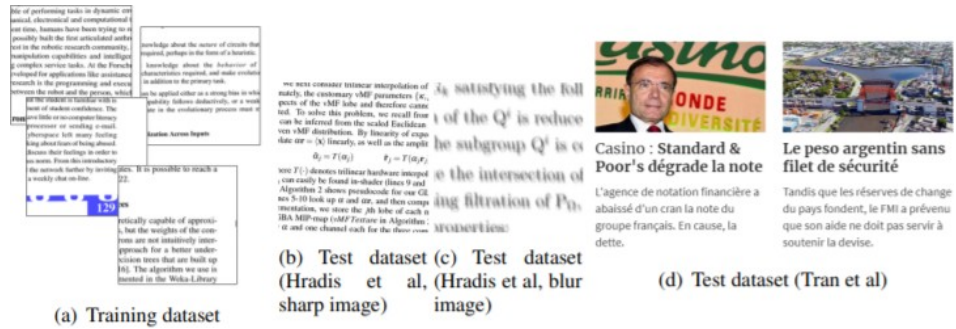

**Fig. 4. Examples of the training and test dataset**

Tran et al. Tran and Ho-Phuoc (2019). It is worth noting that the dataset of Hradis et al. combines two types of blur: motion blur similar to camera shake and defocus blur. Figure 4 shows some representative examples of the training and testing datasets.

*4.2. Quantitative Evaluation*

We have evaluated quantitatively our method using four metrics: the Peak Signal to Noise Ratio (PSNR) Wang and Bovik (2009), the Structural Similarity Index (SSIM) Wang et al. (2004), the Information Fidelity Criterion (IFC) Sheikh et al. (2005), and the Visual Information Fidelity (VIF) Sheikh and Bovik (2006). These metrics are commonly used for image quality assessment in the context of evaluating the performance of text image restoration algorithms. Although these metrics require access to the full reference (the ground truth), this is not a problem with our testing dataset. Moreover, they allow to analyze the quality from four different perspectives: the sensitivity to errors through PSNR (norm of the arithmetic difference between the reference and the output image); the loss of the image structure through SSIM; the resemblance of the statistical information maintained in the output image with respect to the reference image through IFC; and the image quality assessment through VIF. Table 2 shows the values obtained for these metrics according to different scales and selection of subsets extracted from the testing dataset. We make a distinction between blurred and sharp images selected from Hradis et al source Hradiš et al. (2015), and the images taken from Tran et al source Tran and Ho-Phuoc (2019).

In addition, we have compared quantitatively our method with the original method of DCSCN Yamanaka et al. (2017), a variant of our proposed method using a *Sigmoid* function as activator for the feature extraction network, and the methods proposed by Xu et al Xu et al. (2017), Tran et al. Tran and Ho-Phuoc (2019) and Mei et al. Mei et al. (2020). For this comparison, we have used the 100 sharp text images selected from Hradis et al. tested dataset Hradiš et al. (2015) and two different scales. As shown in table 3, our models achieve achieve the best values of SSIM, PSNR IFC and VIF. In the case of scale 4, ST-DCSCN is a bit higher than SDT-DCSCN because the model was trained with sharp text images. Our metric results are better than the results obtained by the other methods.

In the table 4, we have compared quantitatively our models with the state of the-art methods using 100 blurred text images from Hradis et al Hradiš et al. (2015). In the case of scale 4, our metric results are better than the ones obtained by the methods of Tran et al. Tran and Ho-Phuoc (2019), Xu et al. Xu et al. (2017), Mei et al. Mei et al. (2020), and Yamanaka et al. Yamanaka et al. (2017). SDT-DCSCN is a bit low than the method of Quan et al. Quan et al. (2020). For scale 2, our metric result is the best values of SSIM, PSNR, VIF and IFC. For the model of Quan et al. Quan et al. (2020) and Xu et al. Xu et al. (2017) is not available to compare with scale 2.

Last, we have also compared the quality of the OCR output obtained from the improved images generated by ST-DCSCN, SDT-DCSCN and DCSCN Yamanaka et al. (2017), which behaves better than the rest of methods compared in table 4. We have selected randomly 10 images from the list of generated images. After applying Tesseract software[1] to obtain OCR from the sharp images in the testing dataset (used as reference text) and the images returned by the methods, we have computed two metrics to compare the characters identified in each pair of corresponding OCR output files. On the one hand, we have computed the average Levenshtein ratio, which is a similarity string ratio based on the Levenshtein edit distance that counts the minimum number of single-character edits (i.e., insertions, deletions, or substitutions) required to change one word into another. On the other hand, we have computed the average cosine similarity of the character frequency vectors associated to each pair of corresponding files. As shown in table 5, our method performs better in scale 4, i.e. the Levenshtein ratio and the cosine similarity value are higher than for scale 2, because the images are blurred and this case provides the best performance of our model to restore the very low resolution and blurred text images.

---

[1] https://pypi.org/project/pytesseract/

It must be noticed that the obtained OCR is not perfect, even for the sharp images in the testing dataset, because we are using cropped images with incomplete words and lines. In the case of character frequency vectors it must be also noted that we have considered only the detection of letters and digits within the range of printable characters in ASCII encoding.

*4.3. Qualitative Evaluation*

Figures 5 and 6 show two examples of the reconstruction provided by our model. It must be noted that the figures also include a visual comparison with respect to the results obtained by the state-of-the-art methods.

On the one hand, Figure 5 is focused on an example of a low-resolution input image. The output of our method is compared visually with the outputs of other two methods specifically designed for super-resolution, the ones proposed by Mei et al. Mei et al. (2020) and Tran et al. Tran and Ho-Phuoc (2019), obtaining a similar quality.

On the other hand, Figure 6 is focused on an example of a blurry input image. Apart from the quantitative improvement shown in table 3, the visual result of our method provides a much clearer and sharper image in comparison with other methods Xu et al. (2017); Yamanaka et al. (2017); Mei et al. (2020); Tran and Ho-Phuoc (2019); Quan et al. (2020). Our model succeeds to reconstruct high resolution clear images from low blurry inputs containing much clearer characters.

## 5. Conclusion

In this work, we have proposed ST-DCSCN and SDT-DCSCN for super-resolving and deblurring text images. These are feasible approaches to generate high-resolution clear images from blurred low-resolution text images. The main contribution of the paper is the adjustment of parameters in the DCSCN architecture Yamanaka et al. (2017), in particular the identification of the required number of layers to provide appropriate results for text images. In addition, we also proposed a specific process for training the network with a combination of blur and sharp images that generates a better model for dealing with low-resolution blurred images.

In addition, we have compared our proposal with four relevant state-of-the-art methods verifying that our approach performs better, not only in terms of quantitative measures, but also taking into account the visual results. Our method works better for very lower and blurry images, specially for scale 4.

Last, we must remark that this proposal can be efficiently implemented. As mentioned in section 3.2, the image reconstruction network of the training phase has a quick response time by using 1x1 CNNs in comparison to other deep-learning methods using 3x3 CNNs.

**Table 2. Quantitative results of ST-DCSCN and SDT-DCSCN on different testing datasets and scales**

| Testing dataset | scale | Method | PSNR | SSIM | VIF | IFC |
|---|---|---|---|---|---|---|
| Hradis et al.[19] blurred | 4 | ST-DCSCN | 17.724 | 0.74 | 0.242 | 1.830 |
| | | SDT-DCSCN | 20.406 | 0.877 | 0.351 | 2.868 |
| | 2 | ST-DCSCN | 16.878 | 0.67 | 0.214 | 1.751 |
| | | SDT-DCSCN | 16.87 | 0.679 | 0.214 | 1.745 |
| Hradis et al.[19] sharp | 4 | ST-DCSCN | 24.927 | 0.96 | 0.515 | 4.585 |
| | | SDT-DCSCN | 21.488 | 0.919 | 0.404 | 3.278 |
| | 2 | ST-DCSCN | 33.091 | 0.99 | 0.705 | 8.66 |
| | | SDT-DCSCN | 33.063 | 0.991 | 0.704 | 8.648 |
| Tran et al. | 4 | ST-DCSCN | 17.580 | 0.77 | 0.220 | 1.464 |
| | | SDT-DCSCN | 17.447 | 0.75 | 0.201 | 1.246 |
| | 2 | ST-DCSCN | 23.472 | 0.92 | 0.450 | 4.240 |
| | | SDT-DCSCN | 23.440 | 0.92 | 0.450 | 4.245 |

**Table 3. Quantitative comparison with state-of-the-art methods using sharp text images selected from Hradis et al Hradisˇ et al. (2015)**

| Algorithm | Scale | PSNR | SSIM | VIF | IFC |
|---|---|---|---|---|---|
| DCSCNYamanaka et al. (2017) | 2 | 22.184 | 0.930 | 0.434 | 4.449 |
| DCSCN with *Sigmoid*Yamanaka et al. (2017) | 2 | 31.657 | 0.989 | 0.677 | 8.323 |
| Tran et al. Tran and Ho-Phuoc (2019) | 2 | 25.783 | 0.966 | 0.532 | 5.958 |
| Mei et al. Mei et al. (2020) | 2 | 26.975 | 0.973 | 0.566 | 6.22 |
| **ST-DCSCN** | 2 | **33.091** | **0.992** | **0.705** | **8.66** |
| **SDT-DCSCN** | 2 | **33.063** | **0.991** | **0.704** | **8.648** |
| DCSCNYamanaka et al. (2017) | 4 | 16.107 | 0.689 | 0.204 | 1.369 |
| DCSCN with *Sigmoid* Yamanaka et al. (2017) | 4 | 20.452 | 0.894 | 0.370 | 3.071 |
| Tran et al. Tran and Ho-Phuoc (2019) | 4 | 16.711 | 0.747 | 0.237 | 1.868 |
| Mei et al. Mei et al. (2020) | 4 | 16.890 | 0.783 | 0.252 | 1.824 |
| MCGAN Xu et al. (2017) | 4 | 21.332 | 0.922 | 0.397 | 3.087 |
| **ST-DCSCN** | 4 | **24.927** | **0.962** | **0.515** | **4.585** |
| **SDT-DCSCN** | 4 | **21.488** | **0.919** | **0.404** | **3.278** |

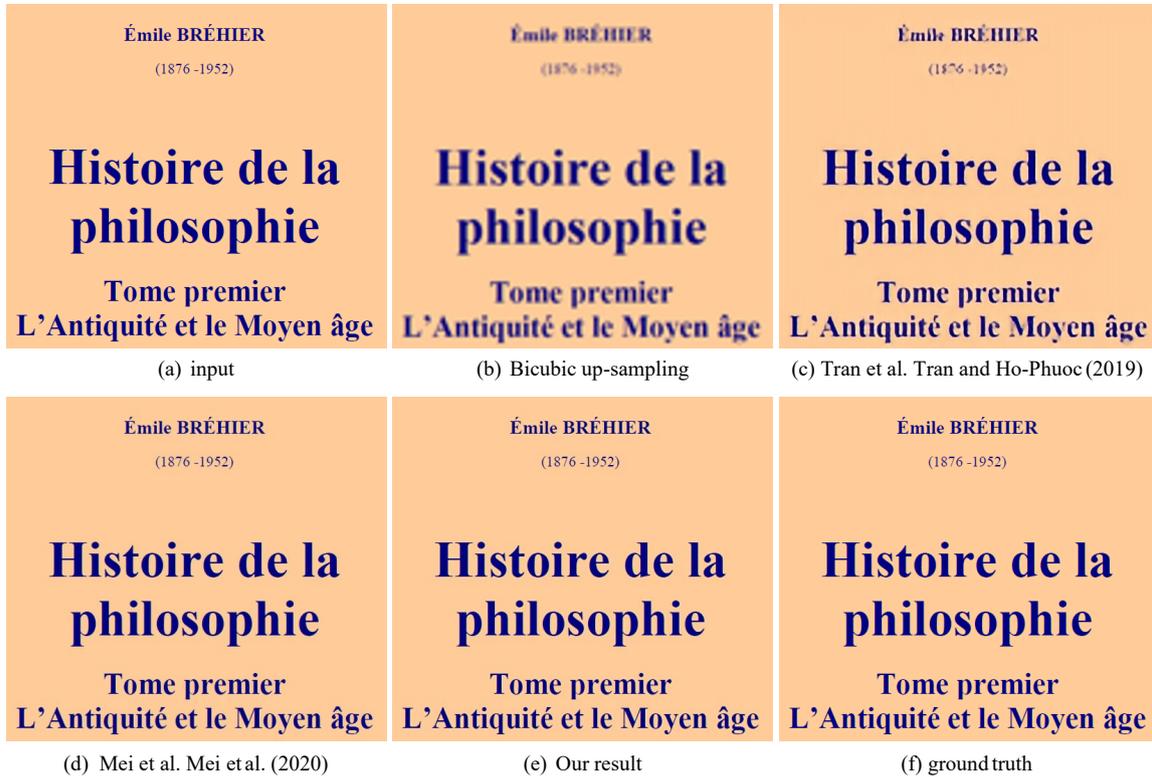

(a) input  (b) Bicubic up-sampling  (c) Tran et al. Tran and Ho-Phuoc (2019)

(d) Mei et al. Mei et al. (2020)  (e) Our result  (f) ground truth

**Fig. 5.** Qualitative comparison of our model with state-of-the-art methods on a low-resolution text image example (scale=2, example taken from Tran et al. test dataset).

Table 4. Quantitative comparison with state-of-the-art methods using blurred text images selected from Hradis et al Hradiš et al. (2015)

| Algorithm | Scale | PSNR | SSIM | VIF | IFC |
|---|---|---|---|---|---|
| DCSCN Yamanaka et al. (2017) | 2 | 16.610 | 0.680 | 0.209 | 1.760 |
| DCSCN with *Sigmoid* Yamanaka et al. (2017) | 2 | 16.984 | 0.685 | 0.216 | 1.735 |
| Tran et al. Tran and Ho-Phuoc (2019) | 2 | 16.440 | 0.665 | 0.201 | 1.648 |
| Mei et al. Mei et al. (2020) | 2 | 16.981 | 0.691 | 0.218 | 1.815 |
| **ST-DCSCN** | 2 | **16.878** | **0.678** | **0.214** | **1.751** |
| **SDT-DCSCN** | 2 | **16.872** | **0.679** | **0.214** | **1.745** |
| DCSCN Yamanaka et al. (2017) | 4 | 15.553 | 0.621 | 0.166 | 1.129 |
| DCSCN with *Sigmoid* Yamanaka et al. (2017) | 4 | 16.734 | 0.732 | 0.214 | 1.641 |
| Tran et al. Tran and Ho-Phuoc (2019) | 4 | 15.313 | 0.656 | 0.159 | 1.045 |
| Mei et al. Mei et al. (2020) | 4 | 15.232 | 0.669 | 0.155 | 1.013 |
| MCGAN Xu et al. (2017) | 4 | 20.128 | 0.897 | 0.364 | 2.828 |
| Quan et al. Quan et al. (2020) | 4 | 25.16 | 0.9695 | 0.514 | 4.355 |
| **ST-DCSCN** | 4 | **17.724** | **0.744** | **0.242** | **1.830** |
| **SDT-DCSCN** | 4 | **20.406** | **0.877** | **0.351** | **2.868** |

Table 5. Quantitative comparison of OCR output using 10 images from Hradis et al dataset Hradiš et al. (2015) blurred version

| Algorithm | Scale | Average cosine similarity | Levenshtein ratio (%) |
|---|---|---|---|
| DCSCN Yamanaka et al. (2017) | 4 | 0.238 | 17.78 |
| ST-DCSCN | 4 | 0.949 | 84.93 |
| SDT-DCSCN | 4 | **0.989** | **92.75** |
| DCSCN Yamanaka et al. (2017) | 2 | 0.832 | 73.23 |
| ST-DCSCN | 2 | 0.781 | 72.72 |
| SDT-DCSCN | 2 | 0.783 | 73.04 |

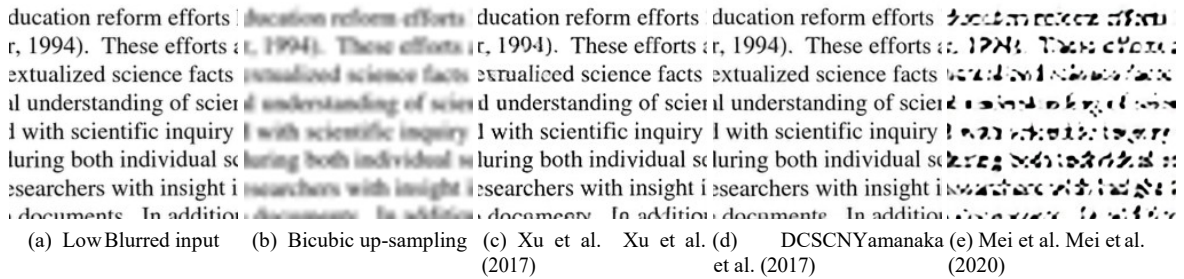

(a) Low Blurred input  (b) Bicubic up-sampling  (c) Xu et al. Xu et al. (2017)  (d) DCSCN Yamanaka et al. (2017)  (e) Mei et al. Mei et al. (2020)

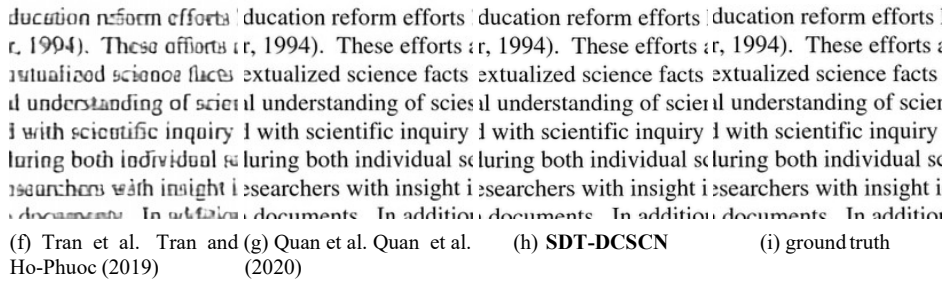

(f) Tran et al. Tran and Ho-Phuoc (2019)  (g) Quan et al. Quan et al. (2020)  (h) **SDT-DCSCN**  (i) ground truth

Fig. 6. Qualitative comparison of our model with state-of-the-art methods on a blurry text image (scale=4, example taken from Hradis et al. test dataset).

## Acknowledgments

The research leading to these results has been partially supported by the Ministry of Higher Education and Scientific Research of Tunisia under the grant agreement number LR11ES48 and , the Spanish Regional Government of Aragon (project T59 20R), and the Spanish Ministry of Science and Innovation (project PID2020-113353RB-I00).We deeply acknowledge Taif University for Supporting this study through Taif University Researchers Supporting Project number (TURSP-2020/327), Taif University, Taif, Saudi Arabia. In addition, we gratefully acknowledge the support of NVIDIA Corporation with the donation of the Quadro P6000 GPU used for this research.

## References

X. Zhang, F. Wang, H. Dong, Y. Guo, A deep encoder-decoder networks for joint deblurring and super-resolution, in: 2018 IEEE International Conference on Acoustics, Speech and Signal Processing (ICASSP), IEEE, 2018a, pp. 1448–1452.

X. Zhang, H. Dong, Z. Hu, W. Lai, F. Wang, M. Yang, Gated fusion network for joint image deblurring and super-resolution, in: British Machine Vision Conference 2018, BMVC 2018, Newcastle, UK, September 3-6, 2018, 153, BMVA Press, 2018b.

Z. Liang, D. Zhang, J. Shao, Jointly Solving Deblurring and Super-Resolution Problems with Dual Supervised Network, in: 2019 IEEE International Conference on Multimedia and Expo (ICME), IEEE, 2019, pp. 790–795.

J. Yamanaka, S. Kuwashima, T. Kurita, Fast and accurate image super resolution by deep CNN with skip connection and network in network, in: International Conference on Neural Information Processing, Springer, 2017, pp. 217–225.

S. Wang, L. Zhang, Y. Liang, Q. Pan, Semi-coupled dictionary learning with applications to image super-resolution and photo-sketch synthesis, in: Proceedings of the IEEE Conference on Computer Vision and Pattern Recognition, 2012, pp. 2216–2223.

J. Yang, J. Wright, T. S. Huang, Y. Ma, Image super-resolution via sparse representation, IEEE transactions on image processing 19 (11) (2010) 2861–2873.

K. Zeng, H. Zheng, Y. Qu, X. Qu, L. Bao, Z. Chen, Single Image Super-Resolution With Learning Iteratively Non-Linear Mapping Between Low- and High-Resolution Sparse Representations, in: 2018 24th International Conference on Pattern Recognition (ICPR), IEEE, 2018, pp. 507–512.

C.-Y. Yang, S. Liu, M.-H. Yang, Structured face hallucination, in: Proceedings of the IEEE Conference on Computer Vision and Pattern Recognition, 2013, pp. 1099–1106.

H. Zhang, J. Yang, Y. Zhang, T. S. Huang, Non-local kernel regression for image and video restoration, in: European Conference on Computer Vision, Springer, 2010, pp. 566–579.

J. Sun, Z. Xu, H.-Y. Shum, Image super-resolution using gradient profile prior, in: 2008 IEEE Conference on Computer Vision and Pattern Recognition, IEEE, 2008, pp. 1–8.

C.-Y. Yang, M.-H. Yang, Fast direct super-resolution by simple functions, in: Proceedings of the IEEE international conference on computer vision, 2013, pp. 561–568.

R. Timofte, V. D. Smet, L. V. Gool, A+: Adjusted anchored neighborhood regression for fast super-resolution, in: Asian conference on computer vision, Springer, 2014, pp. 111–126.

K. Jiang, Z. Wang, P. Yi, J. Jiang, Hierarchical dense recursive network for image super-resolution, Pattern Recognition 107, Article 107475 (2020).

H. T. Tran, T. Ho-Phuoc, Deep Laplacian Pyramid Network for Text Images Super-Resolution, in: 2019 IEEE-RIVF International Conference on Computing and Communication Technologies (RIVF), IEEE, 2019, pp. 1–6.

Y. Mei, Y. Fan, Y. Zhou, L. Huang, T. S. Huang, H. Shi, Image super-resolution with cross-scale non-local attention and exhaustive self-exemplars mining, in: Proceedings of the IEEE/CVF Conference on Computer Vision and Pattern Recognition, 2020, pp. 5690–5699.

M. Ljubenović, L. Zhuang, M. A. Figueiredo, Class-adapted blind deblurring of document images, in: 2017 14th IAPR International Conference on Document Analysis and Recognition (ICDAR), volume 1, IEEE, 2017, pp. 721–726.

X. Jiang, H. Yao, S. Zhao, Text image deblurring via two-tone prior, Neurocomputing 242 (2017) 1–14.

J. Pan, D. Sun, H. Pfister, M.-H. Yang, Blind image deblurring using dark channel prior, in: Proceedings of the IEEE Conference on Computer Vision and Pattern Recognition, 2016, pp. 1628–1636.

M. Hradiš, J. Kotera, P. Zemčík, F. Šroubek, Convolutional neural networks for direct text deblurring, in: Proceedings of BMVC, volume 10, 2015, p. 2.

H. Neji, M. B. Halima, T. M. Hamdani, J. Nogueras-Iso, A. M. Alimi, Blur2sharp: A gan-based model for document image deblurring., Int. J. Comput. Intell. Syst. 14 (2021) 1315–1321.

B. Du, X. Ren, J. Ren, CNN-based Image Super-Resolution and Deblurring, in: Proceedings of the 2019 International Conference on Video, Signal and Image Processing, 2019, pp. 70–74.

H. Liu, J. Qin, Z. Fu, X. Li, J. Han, Fast simultaneous image super-resolution and motion deblurring with decoupled cooperative learning, Journal of Real-time Image Processing 17 (2020) 1787–1800.

J. S. Lumentut, I. K. Park, Deep neural network for joint light field deblurring and super-resolution, in: International Workshop on Advanced Imaging Technology (IWAIT) 2020, volume 11515, International Society for Optics and Photonics, 2020, p. 1151507.

F. Albluwi, V. A. Krylov, R. Dahyot, Super-resolution on degraded low-resolution images using convolutional neural networks, in: 2019 27th European Signal Processing Conference (EUSIPCO), IEEE, 2019, pp. 1–5.

Y. Quan, J. Yang, Y. Chen, Y. Xu, H. Ji, Collaborative deep learning for super-resolving blurry text images, IEEE Transactions on Computational Imaging 6 (2020) 778–790.

J. U. Yun, B. Jo, I. K. Park, Joint face super-resolution and deblurring using generative adversarial network, IEEE Access 8 (2020) 159661–159671.

Y. Li, Z. Yang, X. Mao, Y. Wang, Q. Li, W. Liu, Y. Wang, GAN with Pixel and Perceptual Regularizations for Photo-Realistic Joint Deblurring and Super-Resolution, in: Computer Graphics International Conference, Springer, 2019, pp. 395–401.

B. Du, X. Ren, S. Chen, J. Ren, D. Cao, Image Super-Resolution and Deblurring Using Generative Adversarial Network, in: Proceedings of the 2019 8th International Conference on Computing and Pattern Recognition, 2019, pp. 266–271.

X. Xu, D. Sun, J. Pan, Y. Zhang, H. Pfister, M.-H. Yang, Learning to super-resolve blurry face and text images, in: Proceedings of the IEEE international conference on computer vision, 2017, pp. 251–260.

M. Lin, Q. Chen, S. Yan, Network in network, arXiv preprint arXiv:1312.4400 (2013).

K. He, X. Zhang, S. Ren, J. Sun, Delving deep into rectifiers: Surpassing human-level performance on imagenet classification, in: Proceedings of the IEEE international conference on computer vision, 2015, pp. 1026–1034.

Z. Wang, A. C. Bovik, Mean squared error: Love it or leave it? a new look at signal fidelity measures, IEEE signal processing magazine 26 (2009) 98–117.

Z. Wang, A. C. Bovik, H. R. Sheikh, E. P. Simoncelli, Image quality assessment: from error visibility to structural similarity, IEEE transactions on image processing 13 (2004) 600–612.

H. R. Sheikh, A. C. Bovik, G. De Veciana, An information fidelity criterion for image quality assessment using natural scene statistics, IEEE Transactions on image processing 14 (2005) 2117–2128.

H. R. Sheikh, A. C. Bovik, Image information and visual quality, IEEE Transactions on image processing 15 (2006) 430–444.